\documentclass[sigconf, nonacm]{acmart}
\AtBeginDocument{%
  }

\setcopyright{acmlicensed}
\copyrightyear{2018}
\acmYear{2018}
\acmDOI{XXXXXXX.XXXXXXX}
\acmConference[Conference acronym 'XX]{Make sure to enter the correct
  conference title from your rights confirmation email}{June 03--05,
  2018}{Woodstock, NY}
\acmISBN{978-1-4503-XXXX-X/2018/06}

\usepackage{algorithmic}
\usepackage{algorithm}
\usepackage{tabularx}
\usepackage{array}
\usepackage{multirow}
\newcolumntype{L}{>{\raggedright\arraybackslash}X}

\newcolumntype{R}{>{\raggedleft\arraybackslash}X}

\newcolumntype{C}{>{\centering\arraybackslash}X}

\begin{document}

\title{Accelerating Transistor-Level Simulation of Integrated Circuits via Equivalence of RC Long-Chain Structures}


\author{Ruibai Tang}
\affiliation{%
  \institution{Tsinghua University}
  \city{Haidian Qu}
  \state{Beijing Shi}
  \country{China}}
\email{trb21@tsinghua.org.cn}

\author{Wenlai Zhao}
\affiliation{%
  \institution{Tsinghua University}
  \city{Haidian Qu}
  \state{Beijing Shi}
  \country{China}}
\email{zhaowenlai@tsinghua.edu.cn}

\renewcommand{\shortauthors}{Ruibai Tang and Wenlai Zhao}

\begin{abstract}
Transistor-level simulation plays a vital role in validating the physical correctness of integrated circuits. However, such simulations are computationally expensive. This paper proposes three novel reduction methods specifically tailored to RC long-chain structures with different scales of time constant. Such structures account for an average of 6.34\% (up to 12\%) of the total nodes in the benchmark circuits. Experimental results demonstrate that our methods yields an average performance improvement of 8.8\% (up to 22\%) on simulating benchmark circuits which include a variety of functional modules such as ALUs, adders, multipliers, SEC/DED checkers, and interrupt controllers, with only 0.7\% relative error.
\end{abstract}


\begin{teaserfigure}
  \includegraphics[width=\textwidth]{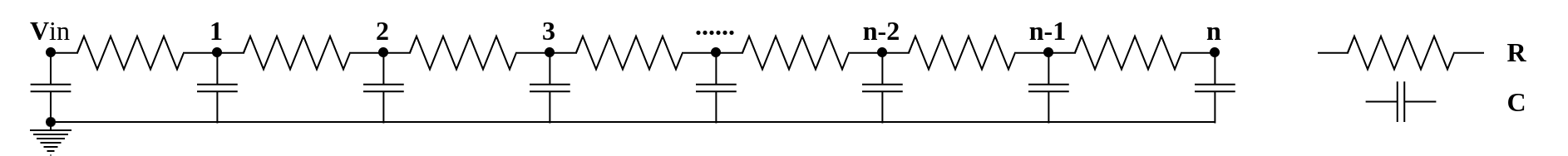}
  \caption{RC Long-Chain Structure with $n$ Identical Resistors and $n+1$ Identical Capacitors }
  \Description{}
  \label{fig:chain_struct}
\end{teaserfigure}


\maketitle

\section{Introduction}

During the actual implementation of hardware, various physical factors, such as semiconductor fabrication, processes and device-level characteristics, which may interfere with circuit functionality. As a result, digital designs on real circuits may fail to operate correctly despite passing Register Transfer Level simulation. To systematically debug such issues, lower-level physical simulations are required. One such approach is \textbf{Transistor-Level Simulation}, which models integrated circuits at a finer granularity, capturing the physical behavior of fundamental components such as resistors, capacitors, and MOSFETs.

Nevertheless, finer-grained simulations demand significantly more computational resources. \texttt{KLU} \cite{davis2010algorithm} is a high-performance sparse matrix solver specifically tailored for circuit simulation. Its paper \cite{davis2010algorithm} demonstrated substantial performance advantages over general-purpose sparse solvers like \texttt{Sparse1.3} \cite{kundert1988user} and \texttt{UMFPACK} \cite{10.1145/992200.992206} in the context of circuit simulation, but experiments presented in \cite{benk2017holistic} reveal that KLU still exhibits significant runtime overhead when applied to large-scale circuits: for the circuits which contain a large number of MOSFETs ($\sim2 \times 10^5$) and relatively few RC components ($\sim10^4$), simulations required approximately \textbf{2400 seconds}; for the circuits characterized by fewer MOSFETs ($\sim 5 \times 10^4$) but extensive RC networks ($\sim 10^6$ components), the simulation time increased drastically to around \textbf{23000 seconds}.

To improve the simulation speed, previous works \cite{810649, 4271561, zyb2024rcreduction, DBLP:journals/corr/abs-0808-4134, chen2012parallel, https://doi.org/10.1002/1099-1506(200010/12)7:7/8<649::AID-NLA217>3.0.CO;2-W, 4796513, kapre2009parallelizing, davis2010algorithm, benk2017holistic, 712097, 384428, 4681658, 7410004, 4253237, 9525071, 20121117, 9963409, 10.1145/3287624.3287658, 8806988, 9358096, 7001355, doi:10.1137/080734029, cgatticer} 
 have tried various optimization techniques. In the domain of circuit reduction—particularly for RC circuits—recent research has increasingly focused on node elimination techniques. \cite{zyb2024rcreduction} Among these, the time-constant balancing reduction method known as \textbf{TICER} \cite{810649, 4271561} represents a leading-edge approach. Extensions of TICER \cite{9525071, cgatticer, 10617546}, as well as its integration with other methods such as spectral sparsification \cite{zyb2024rcreduction}, have significantly improved the efficiency of circuit simplification. Within an acceptable error tolerance, these methods can greatly reduce the complexity of RC circuits and enhance simulation performance. However, such improvements are typically based on the assumption that the eliminated nodes possess \textbf{small time constants} \cite{4271561}. When nodes have large time constants, eliminating them can lead to uncontrolled approximation errors.

Through our case studies on benchmark circuits, we observed the presence of RC long-chain structures, as illustrated in Figure \ref{fig:chain_struct}. On average, such structures account for an average of \textbf{6.34\%} (up to \textbf{12\%}) of the total nodes in the circuits. Although these nodes typically exhibit relatively small time constants, satisfying the assumptions required by TICER,  the reduction methods based on TICER is not that intuitive. In this paper, we propose three direct reduction methods tailored to RC long-chain structures even with the relatively \textbf{large time constants}, which is not able for reduction methods based on TICER to handle.

The contributions of this paper are summarized as follows:

\begin{itemize}
    \item[(1)] We solve the differential equations of RC long-chain structures via Fourier transform to obtain an explicit solution in the frequency domain. Based on numerical insights, we propose three direct reduction methods to handle with different relative scales of time constants.
    \item[(2)] We analyze the errors and the numerical stability of our methods. We also discuss subtle but critical issues such as the causal effect of the impulse response function and the appropriate selection of voltage functions, which are often overlooked and prone to misuse.
    \item[(3)] Experimental results demonstrate that our method yields an average performance improvement of \textbf{8.8\%} (up to \textbf{22\%}) on simulating benchmark circuits which include a variety of functional modules such as ALUs, adders, multipliers, SEC/DED checkers, and interrupt controllers, with only \textbf{0.7\%} relative error.
\end{itemize}

As circuit complexity and density continue to grow rapidly in industrial design. These simulations can take \textbf{several weeks or even months} to complete \cite{20121117}, making any optimization in solver performance potentially impactful, which may reduce simulation time by \textbf{days or even weeks}.

\section{Background and Motivation}

\subsection{Transient Analysis using \texttt{SPICE}}

\texttt{SPICE} \cite{Nagel:M382} (Simulation Program with Integrated Circuit Emphasis) is an algorithm originally developed at the University of California, Berkeley, and is widely used for transistor-level circuit simulation. Algorithm \ref{alg:typical_spice} illustrates a typical workflow of the \texttt{SPICE} algorithm for performing transient simulations at the transistor level.

\begin{algorithm}[h]
    \caption{Typical \texttt{SPICE} for Transient Simulation}
    \label{alg:typical_spice}
    \renewcommand{\algorithmicrequire}{\textbf{Input:}}
    \renewcommand{\algorithmicensure}{\textbf{Output:}}
    
    \begin{algorithmic}[1]
        \REQUIRE $\texttt{.net/.sp/.cir} \text{ Files}$ %%input
        \ENSURE $\text{Simulation } \texttt{Results}$ %%output
            \STATE $\texttt{Circuit} \leftarrow \texttt{Read(.net/.sp/.cir} \text{ File}\texttt{)}$
            \STATE $\text{Pre-processing } \texttt{Circuit}$
            \STATE $t\leftarrow \text{start\_t}$
            \WHILE{$t\le \text{end\_t}$}
                \STATE $\texttt{Matrix} \leftarrow \texttt{Circuit} \text{ State Description (Numerically)}$ 
                \FOR{$i = 0,1, ..., \texttt{max\_iter} $}
                    \STATE $\text{Newton Iteration Settings}$
                    \STATE $\text{Solving Sparse Linear System } \mathbf{Ax}=\mathbf{b}$
                    \STATE $\text{Newton Iteration Updates}$
                \ENDFOR
                \STATE $\text{Recording }\texttt{Results}$
                \STATE $t \leftarrow t+\text{time\_step}$
            \ENDWHILE
    \end{algorithmic}
\end{algorithm}

At each time step, the state of the circuit is abstracted into a system of equations. For pure resistance and capacitance networks, this system can be described by Kirchhoff’s current or voltage laws, allowing for direct solution. 

For example, suppose the conductance matrix of the circuit is denoted by $\mathbf{G}$, the capacitance matrix by $\mathbf{C}$, the external incentives by $\mathbf{b}$ and let $\mathbf{x}$ represent the column vector of node voltages. Then, the circuit state can be described by the differential-algebraic equation:
\begin{align*}
\mathbf{Gx} + \mathbf{C} \frac{{\rm d}\mathbf{x}}{{\rm d}t} = \mathbf{b}
\end{align*}

In \texttt{SPICE} transient simulation, the time step represents the minimum granularity of voltage state updates. Since the derivative term ${\rm d}\mathbf{x}/{\rm d}t$ cannot be computed analytically during simulation, it is approximated using numerical methods. Assuming the current time is $t$ and the time step is $\Delta t$, the derivative is usually approximated as:
\begin{align*}
\frac{{\rm d}\mathbf{x}}{{\rm d}t} \approx \frac{\mathbf{x}(t) - \mathbf{x}(t - \Delta t)}{\Delta t} = \frac{\Delta \mathbf{x}}{\Delta t}
\end{align*}

Therefore, we only need to solve the following linear system, but must be careful to check the numerical stability because the $\Delta t$ is very small:
\begin{align*}
    (\mathbf{G} + \frac{\mathbf{C}}{\Delta t})\Delta\mathbf{x}  = \mathbf{b} - \mathbf{G}\mathbf{x}(t-\Delta t)
\end{align*}

For circuits containing nonlinear components such as semiconductor devices, the resulting system is nonlinear and must be solved using Newton's iterative method.

The nonlinear system to be solved can be formulated as:
\begin{align*}
\mathbf{F}(\mathbf{x}) = 0, \quad \mathbf{F}: \mathbb{R}^n \rightarrow \mathbb{R}^n
\end{align*}

The basic Newton iteration for solving this system is given by:
\begin{align*}
\mathbf{J_F}(\mathbf{x}^{(k)}) \Delta\mathbf{x}^{(k)} = -\mathbf{F}(\mathbf{x}^{(k)}), \quad \mathbf{x}^{(k+1)} = \mathbf{x}^{(k)} + \Delta\mathbf{x}^{(k)}
\end{align*}

Here, $\mathbf{J_F}(\mathbf{x}^{(k)})$ denotes the Jacobian matrix of $\mathbf{F}(\mathbf{x})$ evaluated at $\mathbf{x} =\mathbf{x}^{(k)}$:
\begin{align*}
\mathbf{J_F}(\mathbf{x}) =
\begin{pmatrix}
\frac{\partial F_1}{\partial x_1} & \cdots & \frac{\partial F_1}{\partial x_n} \\
\vdots & \ddots & \vdots \\
\frac{\partial F_n}{\partial x_1} & \cdots & \frac{\partial F_n}{\partial x_n}
\end{pmatrix}
\end{align*}

Upon the completion of each Newton iteration, the circuit state at that time point is determined. Since the simulation spans a time interval, Newton iterations must be carried out at every discrete time point. In general, although the structure of the state equations remains unchanged throughout the simulation, the numerical values vary over time. Therefore, the structural form of the system matrix can be preprocessed prior to the simulation.

In summary, the \texttt{SPICE} algorithm proceeds as follows: it begins by preprocessing the circuit netlist and the symbolic form of the system matrix. A transient time step size $s$ is selected, and the algorithm enters an outer loop that iterates over time steps. Within each time step, an inner loop performs Newton iterations by solving a linear system of the form $\mathbf{Ax} = \mathbf{b}$. As the matrix $\mathbf{A}$ is typically sparse, any suitable sparse linear solver can be employed, with the \texttt{KLU} \cite{davis2010algorithm} algorithm being a common choice.
\begin{table*}[t]
    \centering
    \caption{Meta Information of Benchmark Circuits}
    \label{tab:ngspice_nme}
    \begin{tabularx}{\textwidth}{llRRRRRRR}
    \toprule
    \textbf{Circuit} & \textbf{Function} & \textbf{File Lines} & $\mathbf{N}$ & $\mathbf{M}$ & $\mathbf{nnz}$ & \textbf{R} & \textbf{C} & \textbf{MOSFETs}\\
    \midrule
    c1355 & 32-bit SEC & 21700 & 19484 & 38183 & 95808 & 8976 & 11294 & 2316 \\
    c1908 & 16-bit SEC/DED & 23211 & 23817 & 50339 & 124461 & 8566 & 12213 & 3374 \\
    c2670 & 12-bit ALU and Controller & 42282 & 40451 & 80329 & 200951 & 17019 & 22093 & 5132 \\
    c3540 & 8-bit ALU & 41872 & 49217 & 106738 & 262642 & 15845 & 22774 & 7383 \\
    c432  & 27-Channel Interrupt Controller & 11405 & 9788  & 18618  & 46987 & 4781 & 5863 & 1102 \\
    c499  & 32-bit SEC & 22224 & 19242 & 36982  & 93164 & 9022 & 11452 & 2252 \\
    c5315 & 9-bit ALU & 71933 & 78940 & 165985 & 410731 & 28647 & 38785 & 11082 \\
    c6288 & 16$\times$16 Multiplier & 57425 & 67044 & 145154 & 357319 & 21474 & 31339 & 10112 \\
    c7552 & 32-bit Adder/Comparator & 4024  & 67694 & 178644 & 424773 & 0 & 11012 & 14942 \\
    c880  & 8-bit ALU & 16734 & 14974 & 28595  & 72103 & 6977 & 8677 & 1750 \\
    \bottomrule
    \end{tabularx}
\end{table*}
\subsection{Motivation}

According to the procedure outlined in Algorithm \ref{alg:typical_spice}, there are three main directions for accelerating transient simulation. (1) The first involves preprocessing the input circuit and optimizing the symbolic structure of the system matrix, as exemplified by circuit reduction techniques. (2) The second focuses on improving the two-level nested iteration process, for example, by adopting dynamic strategies that can accelerate convergence or reduce matrix size. (3) The third targets optimization of the sparse matrix solving itself.

Through practical simulation runs and profiling of algorithmic bottlenecks, we find that the majority of simulation time is spent on sparse matrix factorization. Since sparse matrix solving have been extensively studied, further optimization in this area has become increasingly difficult. On the other hand, even with parallelization \cite{chen2012parallel, https://doi.org/10.1002/1099-1506(200010/12)7:7/8<649::AID-NLA217>3.0.CO;2-W, 4796513} and hardware acceleration efforts \cite{kapre2009parallelizing}, \texttt{SPICE}-based methods largely remain constrained within the framework of general-purpose sparse matrix solving optimization. 

In fact, general-purpose optimization often underperform in circuit simulation because they overlook problem-specific characteristics. For instance, while the \texttt{Sparse1.3} \cite{kundert1988user} algorithm is robust for general sparse systems, it exhibits poor performance. In contrast, the \texttt{KLU} algorithm achieves high efficiency by leveraging the specific mathematical properties of circuit matrices—particularly their low $ \text{Flops}/|{\rm L}+{\rm U}| $ ratio. \cite{davis2010algorithm} Although these properties may be implicit and difficult to explain from a structural or physical perspective, they are crucial to \texttt{KLU}’s performance advantage on circuit benchmarks.

Therefore, our work emphasizes the analysis of structural peculiarities in benchmark circuits, and leverages their special properties to improve simulation speed, rather than attempting to perform general-purpose optimization.

\section{Benchmark and Case Study}

We select 10 digital logic circuits originated from the ISCAS'85 benchmark \cite{iscas85} for analysis. As their scales have been expanded in subsequent studies \cite{xujingye}, we adopt these enlarged versions for our experiments. All complete circuit files used in the experiments can be found in the \texttt{examples/klu} directory of the open-source simulator \texttt{Ngspice} \cite{ngspice}. For the original versions of these benchmarks, their RTL-level structures are thoroughly revealed in \cite{10.1109/54.785838}.

\subsection{Meta Information of Benchmark}

Table \ref{tab:ngspice_nme} lists the circuit name, function, number of lines in the netlist file, number of nodes $N$, number of edges $M$, the number of non-zero elements ($nnz$) in the sparse matrix used in \texttt{Ngspice} simulations, and the number of resistors (R), capacitors (C) as well as metal–oxide–semiconductor field-effect transistors (MOSFETs) used in the circuit.

Note that the \texttt{Ngspice} circuit graph modeling algorithm may introduce custom auxiliary nodes, so the number of nodes $N$ is generally greater than or equal to the number of nodes explicitly specified in the netlist. In the sparse matrix derived from the benchmark circuits, if $a \ne b$, then both $(a,b)$ and $(b,a)$ either exist or do not exist simultaneously. However, due to the presence of nonlinear components with asymmetric bidirectional conductance, these two elements are not necessarily numerically equal. In Table \ref{tab:ngspice_nme}, $(a,b)$ is counted as an undirected edge with potentially unequal weights in both directions, and $M$ is determined accordingly.

From a functional perspective, these benchmarks cover several critical categories of digital logic circuits, including arithmetic logic units (ALUs), adders, multipliers, comparators, Hamming code circuits (SEC/DED), interrupt controllers, and controllers. These are widely used in real-world systems: ALUs and controllers are essential components of CPUs; Hamming code-based error correction is frequently implemented in hardware routers; adders, multipliers, and comparators are fundamental building blocks in GPUs, where they are massively integrated for high-throughput computing.

Some circuits within the benchmarks share similar functionality but differ in structural implementation or component usage. This diversity reflects real-world practice, where different vendors may design functionally equivalent circuits with varying architectures due to the lack of standardization. The benchmark suite accounts for this variability by including, for instance, four different ALUs and three implementations of SEC logic, enabling a more comprehensive investigation into common patterns across circuit types.

\begin{table*}[t]
  \centering
  \caption{Resistance-Quantity of Benchmark Circuits}
  \label{tab:resistance_summary}
  \begin{tabularx}{0.95\textwidth}{CcccccccccccC}
    \toprule
    & \textbf{R ($\Omega$)} & \textbf{c1355} & \textbf{c1908} & \textbf{c2670} & \textbf{c3540} & \textbf{c432} & \textbf{c499} & \textbf{c5315} & \textbf{c6288} & \textbf{c7552} & \textbf{c880} &\\
\midrule
& 0.635544 & - & - & - & - & 4748 & - & - & - & - & - &\\
& 0.953316 & 8966 & - & - & - & - & 9022 & - & - & - & 6977 &\\ 
& 1.271088 & - & 8566 & 17019 & - & - & - & - & - & - & -  &\\
& 1.58886 & - & - & - & 15845 & - & - & - & 21474 & - & - &\\
& 1.906632 & - & - & - & - & - & - & 28647 & - & - & - &\\
& $10^6$ & 10 & - & - & - & 33 & - & - & - & - & - &\\
  \bottomrule
\end{tabularx}
\end{table*}

\begin{figure*}
    \centering
    \includegraphics[width=\textwidth]{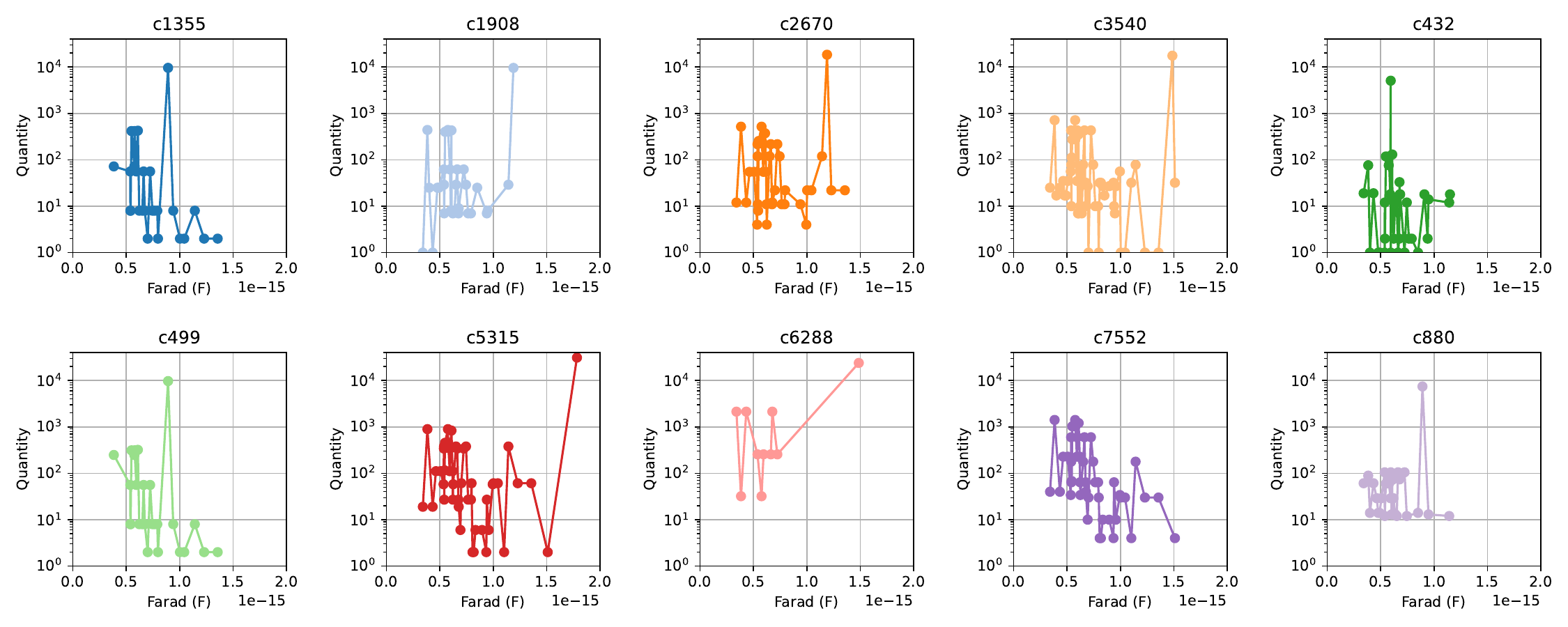}
    \caption{Capacitance-Quantity Distribution of Benchmark Circuits}
    \label{fig:devC_grid}
\end{figure*}

\subsection{Types of Circuit Components}

An inspection of all components in the benchmark circuits reveals that only three types are used: resistors (R), capacitors (C), and metal–oxide–semiconductor field-effect transistors (MOSFETs).

For resistors, Table \ref{tab:resistance_summary} summarizes six distinct resistance values used across the benchmark circuits, along with the number of resistors with each value in each circuit. It can be observed that, except for circuits \texttt{c1355} and \texttt{c432}, which contain a small number of resistors with a resistance of $10^6\Omega$, all other circuits use resistors with a single uniform resistance value. Notably, circuit \texttt{c7552} does not include any resistors.

In contrast, capacitors exhibit much greater variability. The capacitance values are distributed within the range from 0 to $2\times 10^{-15}$F. Figure \ref{fig:devC_grid} shows line charts plotting capacitance versus the number of capacitors for all benchmark circuits. Due to the significant disparity in the number of capacitors for different capacitance values—some capacitance values correspond to tens of thousands of components, while others are represented only sparsely—the vertical axis is log-scaled. 

It is evident that, with the exception of \texttt{c7552}, each circuit has a dominant capacitance value, whose count reaches approximately $10^4$, while the total number of capacitors with all non-dominant values is smaller than that of the dominant group. In contrast, \texttt{c7552} contains multiple capacitance values, each with a quantity on the order of $10^3$, and does not exhibit a clearly dominant capacitance value.

\subsection{Long-chain Characteristics of Circuits}

\begin{figure}[h]
  \centering
  \includegraphics[width=\linewidth]{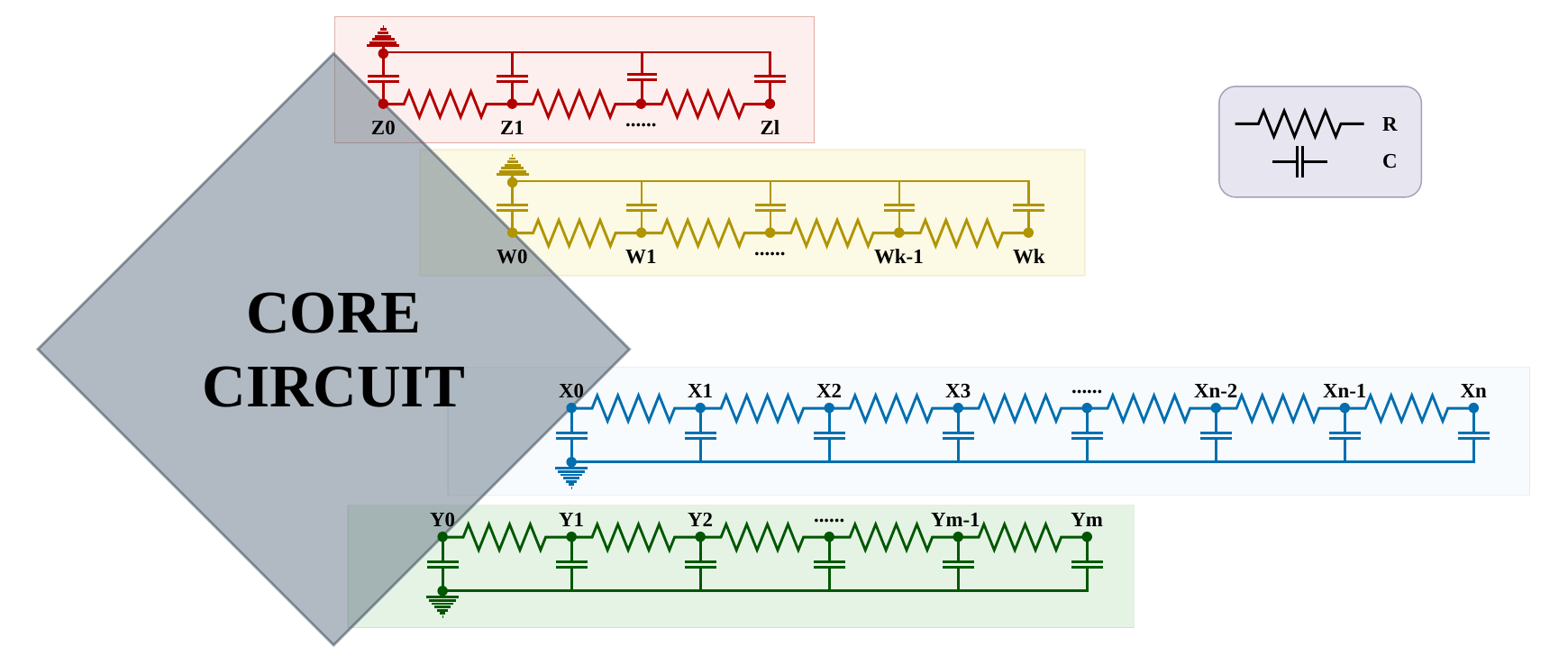}
  \caption{Structural Decomposition of Circuit}
  \label{fig:cir_nuclear}
\end{figure}

\begin{table}
    \centering
    \caption{Maximum Length and Total Number of Nodes in Long Chains and Their Proportion of Total Nodes (MOSFETs Internal Custom Auxiliary Nodes not Included).}
    \begin{tabular}{lcccc}
    \toprule
    \textbf{Circuit} & $\mathbf{N}$ & \textbf{Maxlen} &$\mathbf{N_{\text{tot}}}$ & \textbf{Split Ratio} \\
    \midrule
    \texttt{c1355} & 14811 & 78 & 1380 & 9.32\% \\
    \texttt{c1908} & 17036 & 104 & 1371 & 8.05\% \\
    \texttt{c2670} & 30030 & 126 & 1870 & 6.23\% \\
    \texttt{c3540} & 34401 & 123 & 987 & 2.87\% \\
    \texttt{c432}  & 7548  & 88 & 181 & 2.40\% \\
    \texttt{c499}  & 14697 & 108 & 1619 & 11.02\% \\
    \texttt{c5315} & 56598 & 138 & 6757 & 11.94\% \\
    \texttt{c6288} & 46788 & 139 & 2260 & 4.83\% \\
    \texttt{c7552} & 37603 & 0 & 0  & 0\% \\
    \texttt{c880}  & 11414 & 95 & 772 & 6.76\% \\
    \midrule
    Ave. & & & & \textbf{6.34\%} \\
    \bottomrule
    \end{tabular}
    \label{tab:chain_sum_ratio}
\end{table}

A structural analysis of circuit \texttt{c1355} was performed using Tarjan's algorithm \cite{doi:10.1137/0201010} and other graph methods. The results reveal that the circuit can be decomposed into a core circuit and several long-chain structures, as illustrated in Figure \ref{fig:cir_nuclear}. In \texttt{c1355}, there are a total of 32 such chains, each containing no more than 78 nodes. Within each chain, adjacent nodes are connected by a single resistor, and every node is connected to ground through a capacitor. Importantly, all resistors in these chains have the same resistance value of $0.953316\Omega$, and all capacitors have the same capacitance of $0.891774\times10^{-15}\mathrm{F}$.

Experimental results show that all benchmark circuits conform to the structural pattern shown in Figure \ref{fig:cir_nuclear}. While the specific resistor and capacitor values vary from circuit to circuit, each individual circuit consistently uses \textbf{only one type of resistor and one type of capacitor} in all of its long chains.

Table \ref{tab:chain_sum_ratio} reports the total number of nodes $N_{\text{tot}}$ in the long chains, as well as their proportion relative to the total number of nodes $N$. We define this proportion as the split ratio, as it reflects the structural division between the core circuit and the extended chain regions. It can be observed that, except for circuits \texttt{c3540} (2.87\%), \texttt{c432} (2.40\%), \texttt{c6288} (4.83\%), and \texttt{c7552} (0\%), the split ratio exceeds 6\% in all other cases, reaching up to approximately 12\% in some circuits.

\subsection{Time Constants of Long-chains}

According to TICER \cite{4271561} and assuming that the capacitance and resistance adjacent to the will-be-eliminated node are $C_1, C_2,..., C_n$ and $R_1, R_2,..., R_m$ respectively, then the time constant can be calculated as follows:
\begin{align*}
    \tau_c = \left({\displaystyle \sum_{i=1}^{n}C_i}\right) /\left({\displaystyle \sum_{i=1}^{m}\frac{1}{R_i}}\right)
\end{align*}

In our Long-chain structures shown in Figure \ref{fig:cir_nuclear}, the nodes not on the leftmost and rightmost sides have the time constant $\tau_c = RC/2$, with $0.63\le R \le 1.91$ and $ 0.341 \le 10^{15}C\le 1.79 $ according to Table \ref{tab:resistance_summary} and Figure \ref{fig:devC_grid}. Therefore, the time constants in our benchmark circuits are very small (on the order of $10^{-15}$).

\section{Equivalence of RC Long-Chain Structures}

The long-chain structure studied in this paper is illustrated in Figure \ref{fig:chain_struct}. All resistors and capacitors in the circuit are of the same type, and the chain consists of a total of $n+1$ nodes. The node denoted as $V_{in}$ is connected to the core circuit, while the rightmost node, indexed as $n$, is a closed terminal. Assuming $s$ is the simulation time step (typically $s = 10^{-12}$ second), and consider the real time interval $t \in [0, s]$, which lies between two adjacent transient points. During this interval, the current and voltage externally applied at the $V_{in}$ node are represented by $I(t)$ and $V_0(t)$, respectively, and the voltage at the $i$-th node is denoted $V_i(t)$. By applying Kirchhoff’s Current Law (KCL) at each node, we obtain a system of differential equations.
\begin{align}
\left\{
\begin{aligned}
 C\frac{{\rm d}V_{i}}{{\rm d}t} + \frac{V_{i}-V_{i-1}}{R} + \frac{V_{i}-V_{i+1}}{R} &= 0, \quad i=1,2,...,n-1 \\
 C\frac{{\rm d}V_n}{{\rm d}t} + \frac{V_{n}-V_{n-1}}{R} &= 0 \\
 C\frac{{\rm d}V_0}{{\rm d}t} + \frac{V_0-V_1}{R} &= I 
\end{aligned}
\right.
\label{diff_eq}
\end{align}

A common approach to solving such systems is to apply the Fourier transform to convert the equations from the time domain $t$ to the frequency domain $\omega$.
\begin{align}
\dot{V}(\omega) &= \int_{-\infty}^{\infty}V(t)\exp( -j \omega t){\rm d}t
\label{fourier}
\end{align}

Differentiate both sides of \eqref{fourier} with respect to $t$.
\begin{align*}
    \frac{{\rm d}}{{\rm d}t}\dot{V}(\omega) &= (- j \omega ) \dot{V}(\omega) + \int_{-\infty}^{\infty}\frac{{\rm d}V}{{\rm d}t}(t) \exp(-j \omega t){\rm d}t
\end{align*}

Note that $\dot{V}(\omega)$ does not depend on $t$, so the left-hand side derivative becomes zero, which leads to a simplified form.
\begin{align}
    j \omega \dot{V}(\omega) &=  \int_{-\infty}^{\infty}\frac{{\rm d}V}{{\rm d}t}(t) \exp(- j \omega t){\rm d}t
    \label{diff_fourier}
\end{align}

Apply the Fourier transform \eqref{fourier} and \eqref{diff_fourier} to \eqref{diff_eq} and rearranging terms yields a new system of algebraic equations. 
\begin{align}
\left\{
\begin{aligned}
 (2+ RC j\omega) \dot{V}_i - \dot{V}_{i-1}  &= \dot{V}_{i+1}, \quad i=1,2,...,n-1 \\
 (1+RC j \omega) \dot{V}_n  &= \dot{V}_{n-1} \\
 (1+RC j \omega) \dot{V}_0 -R\dot{I}  &= \dot{V}_1 
\end{aligned}
\right.
\label{alg_eqs}
\end{align}

The first equation in \eqref{alg_eqs} is a second-order linear recurrence relation, whose general solution can be obtained via the method of characteristic roots.
\begin{align}
\dot{V}_k = C_0a^k+C_1b^{k}, \quad k = 0,1,...,n
\label{general_solution_of_Vk}
\end{align}

where
\begin{align*}
a &= 1+ RC j \omega /2+ \sqrt{-R^2C^2\omega^2/4+ RCj\omega} \\
b &= 1+ RC j \omega/2 - \sqrt{-R^2C^2\omega^2/4+ RCj\omega}
\end{align*}

Substitute $k = 0, 1$ in \eqref{general_solution_of_Vk}, and use the third equation in \eqref{alg_eqs}, the constants $C_0$ and $C_1$ are determined by the boundary conditions $\dot{V}_0$ and $\dot{I}$.
\begin{align*}
\left\{
\begin{aligned}
C_0 + C_1 &= \dot{V}_0  \\
C_0 a+C_1b &= (1+ RC j\omega)\dot{V}_0 - R\dot{I}  
\end{aligned}
\right.
\end{align*}

Let $\tau = 1 + RC j\omega$, we derive the explicit solution.
\begin{align*}
\left\{
\begin{aligned}
C_0 &= \frac{(b-\tau)\dot{V}_0 +R\dot{I}}{b-a}  \\
C_1 &= \frac{(\tau-a)\dot{V}_0 -R\dot{I}}{b-a}
\end{aligned}
\right.
\end{align*}

Combining \eqref{general_solution_of_Vk} with the second equation in \eqref{alg_eqs}, we can establish the relationship between $\dot{V}_0$ and $\dot{I}$.
\begin{align}
    \dot{I} &= \frac{\dot{V}_0}{R}\frac{(\tau-b)(\tau a^n -a^{n-1})+(a-\tau)(\tau b^n - b^{n-1})}{\tau a^n- a^{n-1}-(\tau b^n - b^{n-1})} \nonumber\\
    & = \dot{V}_0 / \dot{Z_n} = \dot{V}_0 \cdot \dot{Y}_n \label{dot_product}
\end{align}

This result indicates that the external current $I$ entering the long-chain structure is completely determined by the port voltage $V_{in}$ and the length of chain $n$.  

Considering the smallness of the time step $s =10^{-12}$, we approximate $V_0(t)$ as a almost linear function over the interval $[0, s]$, and extend the definition of the linear function over the whole $\mathbb{R}$ for convergent Fourier transform by multiplying $e^{-M^2t^2}$ with $M\in[1,10^{-6}/s]$, which helps us satisfy all the integral convergence conditions in this paper. Let $ \hat{v} = V_0(s), v = V_{0}(0), \Delta v = \hat{v} -v$, we have
\begin{align}
    V_{0}(t) = \left(\frac{\Delta v}{s}t+v\right)e^{-M^2t^2}
    \label{approximate_v0}
\end{align}

And its Fourier transform follows.
\begin{align*}
    \dot{V}_0(\omega)=-j\frac{\sqrt{\pi}}{2M^3}\omega e^{-{\omega}^2/(4M^2)} \frac{\Delta v}{s} + \frac{\sqrt{\pi}}{M}e^{-\omega ^2/(4M^2)}v
\end{align*}

Let 
\begin{align*}
    \dot{I}(\omega) = \dot{V}_0(\omega)\cdot \dot{Y}_n(\omega) = \dot{F}_n(\omega) \frac{\Delta v}{s} + \dot{G}_n(\omega) v
\end{align*}

Apply the inverse Fourier transform at $t= s$, we obtain the following expression.
\begin{align}
    {I}(s) &=F_n(s) \frac{\Delta{v}}{s} + G_n(s) v 
    \label{final_linear_expression}
\end{align}

where 
\begin{align*}
    F_{n}(s) 
    &= \frac{-j}{4\sqrt\pi M^3}\int_{-\infty}^{\infty}\omega \dot{Y}_n(\omega)\exp\left(-\omega^2/(4M^2)+j\omega s\right){\rm d}\omega\\
    G_{n}(s) 
    &= \frac{1}{2\sqrt\pi M}\int_{-\infty}^{\infty}\dot{Y}_n(\omega)\exp\left(-\omega^2/(4M^2)+j\omega s\right){\rm d}\omega
\end{align*}

\subsection{Implementation Details}

\subsubsection{Causal Effect of Impulse Response $Y_{n}(t)$}

\begin{align}
    I(t) =  (V_0 *Y_n)(t) = \int_{-\infty}^{\infty}V_0(\tau)Y_{n}(t-\tau){\rm d}\tau \label{incorrect_i_expr}
\end{align}

Apply the inverse Fourier transform on equation \eqref{dot_product} at $t$, we obtain the expression \eqref{incorrect_i_expr}.

However, the impulse response function $Y_n(t-\tau)$ has non-zero values when $t-\tau<0$, which does not follow the causal effect and makes no practical significance. That is, $I(t)$ should not depend on the voltage $V_0(\tau)$ at a later time point $\tau\,(t< \tau) $. Therefore, the theoretically correct expression should have the upper limit of the integral as shown below
\begin{align}
I_\text{real}(t) = \int_{-\infty}^{t} V_0(\tau)Y_n(t-\tau){\rm d}\tau \label{real_conv_expr_ofi}
\end{align}

Of course, if we insist on using the form of equation~\eqref{incorrect_i_expr}—for instance, to simplify some derivation in the frequency domain—there are two alternative strategies to address the resulting issues.

First, we can define $V_0(t, \infty) := 0$. For example, We use $e^{-M^2 t^2}$ in equation \eqref{approximate_v0}, which ensures that $V_0(t)$ decays rapidly to zero for $t > s$. In this case, no modification of the frequency response $\dot{Y}_n(\omega)$ is required to enforce causality.

Second, we can directly modify $\dot{Y}_n(\omega)$. We define the real inpulse response function as $Y_{n,\text{real}}(t) := Y_n(t)\, \theta(t)$, where $\theta(t)$ denotes the unit step function: $\theta(t)=0$, $t<0$ and $\theta(t)=1$, $t>0$.

Taking the Fourier transform of $Y_{n,\text{real}}(t)$, we obtain:
\begin{align*}
    \dot{Y}_{n,\text{real}}(\omega) &= \int_{-\infty}^{\infty} \dot{Y}_n(\omega - \nu)\, \dot{\theta}(\nu)\, {\rm d}\nu \\
    &= \int_{-\infty}^{\infty} \dot{Y}_n(\omega - \nu) \left( \pi \delta(\nu) + \frac{1}{j\nu} \right) {\rm d}\nu \\
    &= \pi \left( \dot{Y}_n(\omega) -j \mathcal{H}[\dot{Y}_n](\omega) \right),
\end{align*}

where $\delta(\nu)$ is the Dirac delta function and $\mathcal{H}$ denotes the Hilbert transform. P.V. denotes the Cauchy principal value.
\begin{equation*}
    \mathcal{H}[f](t) = \frac{1}{\pi} \, \text{P.V.} \int_{-\infty}^{\infty} \frac{f(\tau)}{t - \tau}\, {\rm d}\tau,
\end{equation*}

\subsubsection{Calculate $\dot{Y}_n(\omega)$, $F_n(s)$, and $G_n(s)$}

\begin{align*}
        \dot{H}_n(\omega) &= \frac{\tau b -1}{\tau a-1}\left(\frac{b}{a}\right)^{n-1} \\
    \dot{Y}_n(\omega) &= \frac{(\tau-b) + (a-\tau)\dot{H}_n(\omega)}{1-\dot{H}_n(\omega)}\frac{1}{R}
\end{align*}

We define $\dot{H}_n(\omega)$ as shows above and it can help us calculate $\dot{Y}_n(\omega)$ in a more numerically stable way. $\dot{H}_n(\omega)$ and $\dot{Y}_n(\omega)$ have the following properties:
\begin{align*}
    \lim_{\omega \rightarrow 0}\dot{H}_n(\omega)&=\lim_{\omega\rightarrow 0}\frac{-\sqrt{RCj\omega/4+1}+O(\omega^{1/2})}{+\sqrt{RCj\omega/4+1}+O(\omega^{1/2})}=-1 \\
    \lim_{\omega\rightarrow\infty}\dot{H}_{n}(\omega)& = \left(\frac{RCj - \sqrt{- R^{2} C^{2}}}{RCj + \sqrt{- R^{2} C^{2}}}\right)^{n - 1} = 0
 \\
    \lim_{\omega \rightarrow0}\frac{\dot{Y}_n(\omega)}{\omega} &= (n+1)Cj, \,\,\,\,\,\,\,\,\,\,\, \lim_{\omega \rightarrow \infty}\frac{\dot{Y}_{n}(\omega)}{\omega}  = Cj
\end{align*}

For $F_n(s), \,G_n(s)$, it's recommended to let $\lambda = \omega /M$, then use the following numerically improved equations for calculation: 
\begin{align*}
    F_{n}(s) 
    &= \frac{-j}{4\sqrt\pi }\int_{-25}^{25}\lambda ^2\frac{\dot{Y}_n(\lambda M)}{\lambda M}\exp\left(-\lambda^2/4+j\lambda (Ms)\right){\rm d}\lambda\\
    G_{n}(s) 
    &= \frac{M}{2\sqrt\pi}\int_{-25}^{25}\lambda \frac{\dot{Y}_n(\lambda M)}{\lambda M}\exp\left(-\lambda^2/4+j\lambda (Ms)\right){\rm d}\lambda
\end{align*}

\subsubsection{Selection of Voltage Function $V_0(t)$}

\begin{align}
    \dot{V}_0(\omega)=\frac{\exp(-j\omega s)(1+j\omega s)-1}{\omega^2}\frac{\Delta v}{s} + \frac{1-\exp(-j\omega s)}{j\omega} v \label{another_approach_to_choose_v}
\end{align}

There is also another way to choose $V_0(t)$ as a linear function $(\Delta v/s)t+v$ when $t\in[0,s]$ with $V_0(t)=0, \, t\in \mathbb{R}\backslash [0,s]$. And its Fourier transform is shown in \eqref{another_approach_to_choose_v}. However, the integrals to calculate $F_n(s), G_n(s)$ do not converge because $\lim_{\omega\rightarrow \infty} \dot{Y}_n(\omega)/\omega = Cj$ and therefore $F_n(s), G_n(s)$ need to integrate $1/\omega$, $\exp(j\omega s)$ in an infinite interval. 
\begin{align*}
    F_{n}(s) 
    &= \frac{1}{2\pi}\int_{-\infty}^{\infty} \frac{(1+j\omega s)-\exp(j\omega s)}{\omega}\frac{\dot{Y}_n(\omega)}{\omega}{\rm d}\omega\\
    G_{n}(s) 
    &= \frac{-j}{2\pi}\int_{-\infty}^{\infty}\left(\exp(j\omega s)-1\right)\frac{\dot{Y}_n(\omega)}{\omega}{\rm d}\omega
\end{align*}

That is also why we finally choose the equation \eqref{approximate_v0} instead of this simpler function.

\subsection{Relatively Large Time Constant}

Table \ref{tab:numerical_values_of_the_first_10} presents the first ten instances of $F_n(s)$ and $G_n(s)$ with parameters $s = 10^{-12}$, $M = 10^4$, $R = 1\Omega$, and $C = 1$F, along with their estimated integral absolute errors. The results show that when the chain grows in length, the current expression almost never changes.
It is a sign that the first two nodes, $V_0$ and $V_1$, in the long-chain structure dominate the current flowing into the system, however, we still have to prove it.

Take a look into the case $n=1$, the voltages of the two nodes $V_0$ and $V_1$ satisfy the following relation:
\begin{align}
    \dot{V}_1(\omega) &= \frac{1}{1 + RCj\omega}\dot{V}_0(\omega)\nonumber \\
    V_1(t)
    &=\frac{1}{RC}\int_{0}^{\infty}V_0(t-\tau)e^{-\tau /RC} {\rm d}\tau \label{V1_expression}
\end{align}

Consequently, the entering current $I(t)$ follows.
\begin{align}
    I(t) &= C\left(\frac{{\rm d}V_0}{{\rm d}t} + \frac{{\rm d}V_1}{{\rm d}t}\right) \nonumber\\ 
    & = C\frac{{\rm d}V_0}{{\rm d}t} + \frac{1}{R}V_0 -  \frac{1}{R^2C}\int_{0}^{\infty}V_0(t-\tau) e^{-\tau /RC} {\rm d}\tau \label{current_math} \\
    &= C\frac{{\rm d}\hat{V}_0}{{\rm d}t} + \frac{1}{R}\hat{V}_0  - \frac{1}{R^2C}\int_{0}^{t+t_0}\hat{V}_0(t-\tau)e^{-\tau/RC}{\rm d}\tau \label{current_v_zero}\\
    \hat{V}_0(t) &= V_0(t)-V_0(-\infty), \, \hat{V}_0(-\infty,-t_0] =0 \nonumber
\end{align}

In typical simulation practice, it is assumed that $V_0$ maintains a constant initial value $V_0(-\infty)$ prior to the start of the simulation. Thus, when the simulation begins at $t=-t_0$, all nodes within the long-chain structure share this same initial voltage $V_0(-\infty)$. Subtracting this offset from $V_0(t)$ does not alter the form of equation \eqref{current_math}, as the constant term $ V_0(-\infty)/R$ is effectively eliminated. This leads to equation \eqref{current_v_zero}.

Since the specific value of $V_0(-\infty)$ is inconsequential for simulation accuracy, we may simply adjust by adding or subtracting this initial offset as needed. This also justifies the assumption $\lim_{t \to -\infty} V_0(t) = 0$ in equation \eqref{approximate_v0}.

When the time constant $\tau_c=RC = 1$ in expression \eqref{V1_expression} is relatively large compared to the elapsed simulation time $d=t+t_0 \approx 10^{-9}$, then we have
\begin{align}
    V_1(t) &=V_0(-\infty) + \frac{1}{\tau_c}\int_0^{d} \hat{V}_0(t-\tau)e^{-\tau/\tau_c}{\rm d}\tau \label{v1_expr_div}\\
    &=V_0(-\infty) + \frac{d}{\tau_c}\mathbb{E}\left[\hat{V}_0[-t_0,t]\right] \nonumber
\end{align}

Considering that $\hat{V}_0$ is usually in the scale of single digit (e.g. $1$V), it shows that there is not enough time for $V_1$ to repond when $V_0$ changes, so $V_1$ almost retains the initial value. That's the meaning of time constant $\tau_c$. Since $V_1$ almost never changes and the later nodes with the same time constant $\tau_c$ depend on $V_1$, it's clear that the influence on their voltages is negligible.

\begin{table}[t]
  \centering
  \caption{$F_n(s),G_n(s)$ with $s=10^{-12}, M =10^4, R=1\Omega, C=1{\rm F}$ and their integral absolute errors }
  \label{tab:numerical_values_of_the_first_10}
  \begin{tabularx}{\linewidth}{lCCCC}
    \toprule
    $\mathbf n\le 10$&$\mathbf F_{n}(s)$&\textbf{Ferr(abs)}& $\mathbf G_n(s)$&\textbf{Gerr(abs)}\\
    \midrule
     1 & 1.0000e+00 & 8.1610e-09 & 9.9980e-01 & 2.5504e-09 \\
  2 & 1.0000e+00 & 8.1610e-09 & 9.9980e-01 & 8.6199e-09 \\
 $\ge3$ & 1.0000e+00 & 8.1610e-09 & 9.9980e-01 & $\le$1.6e-08 \\
  \bottomrule
\end{tabularx}
\end{table}

\subsection{Relatively Small Time Constant}

Table \ref{tab:numerical_values_of_the_first_10_small_time_constant} presents the first ten instances of $F_n(s)$ and $G_n(s)$ with parameters $s = 10^{-12}$, $M = 10^4$, $R = 1\Omega$, and $C = 10^{-15}$F, along with their estimated integral absolute errors. The results show that the current $I$ is almost $(n+1)C$ times of $\Delta v/s \approx {\rm d}V_0/{\rm d}t$.
It is a sign that the voltages $V_i$ of all nodes in the chain change simultaneously and uniformly as $V_0$ does. We will give a gist proof.

We still first take a look into the case $n=1$, the equation \eqref{V1_expression} is equivalent to 
\begin{align}
     V_1(t) &= V_0(-\infty)+\int_0^{(t+t_0)/RC} \hat{V}_0(t-RC\tau)e^{-\tau}{\rm d}\tau \nonumber \\
     &= V_0(-\infty)+\int_0^{s/RC} \hat{V}_0(t-RC\tau)e^{-\tau}{\rm d}\tau \label{v1_expr_small_time_constant} 
\end{align}

When the time constant $\tau_c=RC=10^{-15}$ is relatively small compared to the simulation time step $s=10^{-12}$ and the elapsed simulation time $d=t+t_0\approx 10^{-9}$, we derive the equation \eqref{v1_expr_small_time_constant} because the error is less than $(d/\tau_c)e^{-s/\tau_c}\mathbb{E}\left|\hat{V}_0[-t_0, t-s]\right|$ which is negligible.

Let $\hat{v}=\hat{V}_0(t), v=\hat{V}_0(t-s), \Delta v=\hat{v}-v$, we can approximate the integral in equation \eqref{v1_expr_small_time_constant} with
\begin{align*}
    &\int_0^{s/RC} \left(\frac{\Delta v}{s}(s-RC\tau) +v\right)e^{-\tau}{\rm d}\tau \\
    = \,& \hat{v}\left(1-e^{-s/\tau_c}\right)+ \Delta v\left(e^{-s/\tau_c}(1+\tau_c/s)-\tau_c/s\right) \\
    = \,& \hat{v} -\tau_c \frac{\Delta v}{s}
\end{align*}

Therefore, we have
\begin{align*}
    V_1(t)=V_0(t) -\tau_c\frac{{\rm d}V_0}{{\rm d}t}
\end{align*}

Since $\left|\tau_c \left({\rm d}V_0 /{\rm d}t\right)\right| \approx |\tau_c\Delta v/s|\le (2\tau_c/s)\sup|V_0|$, that's why $V_0, V_1$ change simultaneously and uniformly, and if we make a simplification that $V_m$ only or to a great extent depends on $V_{m-1}$, then we have
\begin{align*}
    V_m(t) &= V_{m-1}(t)-\tau_c\frac{{\rm d}V_{m-1}}{{\rm d}t} \\
    &= V_0(t) +\sum_{k=1}^{m}(-\tau_c)^{k} \binom{m}{k}\frac{{\rm d}^kV_0}{{\rm d}t^{k}}
\end{align*}

Now, we calculate the current $I(t)$
\begin{align*}
    I(t) &= C\sum_{m=0}^{n}\frac{{\rm d}V_m}{{\rm d}t} \\
    & =(n+1)C\frac{{\rm d}V_0}{{\rm d}t} + \sum_{m=1}^{n}\sum_{k=1}^{m}(-\tau_c)^{k} \binom{m}{k}C\frac{{\rm d}^{k+1}V_0}{{\rm d}t^{k+1}} \\
    & = (n+1)C\frac{{\rm d}V_0}{{\rm d}t} - \frac{n(n+1)}{2}RC^2\frac{{\rm d}^2V_0}{{\rm d}t^2} + o\left(\left(\frac{\tau_c}{s}\right)^2\right)
\end{align*}

\begin{table}[t]
  \centering
  \caption{$F_n(s),G_n(s)$ with $s=10^{-12}, M =10^4, R=1\Omega, C=10^{-15}{\rm F}$ and their integral absolute errors }
  \label{tab:numerical_values_of_the_first_10_small_time_constant}
  \begin{tabularx}{\linewidth}{lCCCC}
    \toprule
    $\mathbf n\le 10$&$\mathbf F_{n}(s)$&\textbf{Ferr(abs)}& $\mathbf G_n(s)$&\textbf{Gerr(abs)}\\
    \midrule
    1 & 2.0000e-15 & 3.2573e-15 & 5.2097e-18 & 1.0377e-17 \\
2 & 3.0000e-15 & 4.8859e-15 & 5.0105e-18 & 1.0245e-17 \\
$\ge 3$ & (n+1)e-15 & $\le$1.7e-14 & $\le$4.9e-18 & $\le$1.0e-17 \\
  \bottomrule
\end{tabularx}
\end{table}

\subsection{Time Constant with Same Order}

This represents the most challenging scenario, where the circuit time constant $\tau_c$ is on the same order of magnitude as the simulation time steps $t$. 

We still begin with the case of $n = 1$ as described by equation~\eqref{V1_expression}. In the equation, the exponential decay term $e^{-\tau / \tau_c}$ does not rapidly diminish, meaning that $V_1(t)$ significantly depends on the values of $V_0$ at earlier time points. Consequently, our previously proposed approximation in equation~\eqref{approximate_v0}, along with its subsequent derivations, becomes invalid. These approximations only account for the contribution from adjacent time steps and fail to incorporate long-range temporal dependencies.

To address this, we turn our attention to directly evaluating the convolution integral in equation~\eqref{real_conv_expr_ofi}, which is derived from the explicit frequency-domain solution and hence provides the theoretically correct result.

Nevertheless, the current approach does not yet support efficient computation of this integral at every $t$. Observing that $V_0(t)$ inherently depends on the history of $I(t - ks)$, we are motivated to approximate the integral using an $m$-order recurrence relation:
\begin{align*}
I(t) &= \int_{-\infty}^{t} V_0(\tau)\, Y_n(t - \tau) \, {\rm d}\tau \\
& =\sum_{k=0}^{m} \frac{I^{(k)}(t-s)}{k!}s^k + o(s^{m})\\
     &\approx \sum_{k=1}^{m} \gamma_k I(t - ks) + \sum_{k=0}^{m} \beta_k V_0(t - ks),
\end{align*}

In practice, the order $m$ is better to meet the condition $m \le n$; otherwise, one might as well solve the original circuit equations directly using LU factorization with lower computational overhead. Furthermore, the coefficients $\gamma_k$ and $\beta_k$ should be pre-computed based on the expected class of input waveform $V_0$, such as smooth sinusoidal signals (\texttt{SIN}) or pulse-like signals (\texttt{PULSE}), to minimize the approximation error.

In this paper, we only propose this approximation strategy as a conceptual direction, without delving into its detailed implementation or accuracy analysis, which we leave as future work.

\subsection{Brief Summary}

Let the simulation start time $-t_0$, the current transient time $t$, the elapsed simulation time $d = t + t_0$, the time step $s$, the time constant $\tau_c = RC$, the input voltage $V_0(t)$ ($V_0(-\infty,-t_0]$ is constant) along with $\hat{V}_0(t) = V_0(t) - V_0(-\infty)$, and parameters $\alpha = 10$, $\gamma_k,\beta_k$. We approximate the inflow current $I(t)$ for the RC long-chain structure using the following criteria:

\begin{itemize}
    \item If $\tau_c < s/\alpha$, then
    \begin{align}
        I(t) = (n+1)C \frac{{\rm d}\hat{V}_0}{{\rm d}t} - \frac{n(n+1)}{2}RC^2\frac{{\rm d}^2 \hat{V}_0}{{\rm d}t^2} \label{quick_node}
    \end{align}
    Note: The condition indicates that all the nodes in the system can immediately respond when $V_0$ changes. When $n\le 64$, we can use \eqref{quick_node} for approximation; but when $n>64$, it's recommended to directly solve the $n' =\lceil\log_2 n\rceil$ or $\lfloor{n/2}\rfloor$ case with equally-shared capacitance of the original system (e.g. for $\lfloor{n/2}\rfloor$ case the capacitance of each node is nearly doubled of the origin) for balanced error and elimination.
    \item If $s/\alpha \le \tau_c \le \alpha \cdot d$, then
    \begin{align}
        I(t) = \sum_{k=1}^m\gamma_kI(t-ks) +\sum_{k=0}^{m} \beta_kV_0(t-ks) \label{case2}
    \end{align}
    Note: In this case, the time constant $\tau_c$ and the simulation steps $t$ are really close, which results in chaos of the system because every node $V_i$ highly depends on the previous curve of $V_0$ and performs convolution with it with almost different weights. It's recommended to properly choose $m\le n$ (otherwise we could directly solve the original linear system by LU factorization in less time) and pre-process the parameters $\gamma_k,\beta_k$ which will lead to mininum error based on your $V_0$ curve type. This situation still needs further study.

    \item If $\tau_c > \alpha \cdot d$, then
    \begin{align}
        I(t)=  C\frac{{\rm d}\hat{V}_0}{{\rm d}t} + \frac{1}{R}\hat{V}_0-\frac{d}{R^2C}\mathbb{E}\left[\hat{V}_0[-t_0, t]\right] \label{case3}
    \end{align}
    Note: This situation implies the leftmost two nodes dominate the system becasue the elapsed time $d$ is too small compared to the time constant $\tau_c$ and there's not enough time for other nodes to respond.
\end{itemize}

\section{Evaluation}

\begin{figure*}
    \centering
    \includegraphics[width=\textwidth]{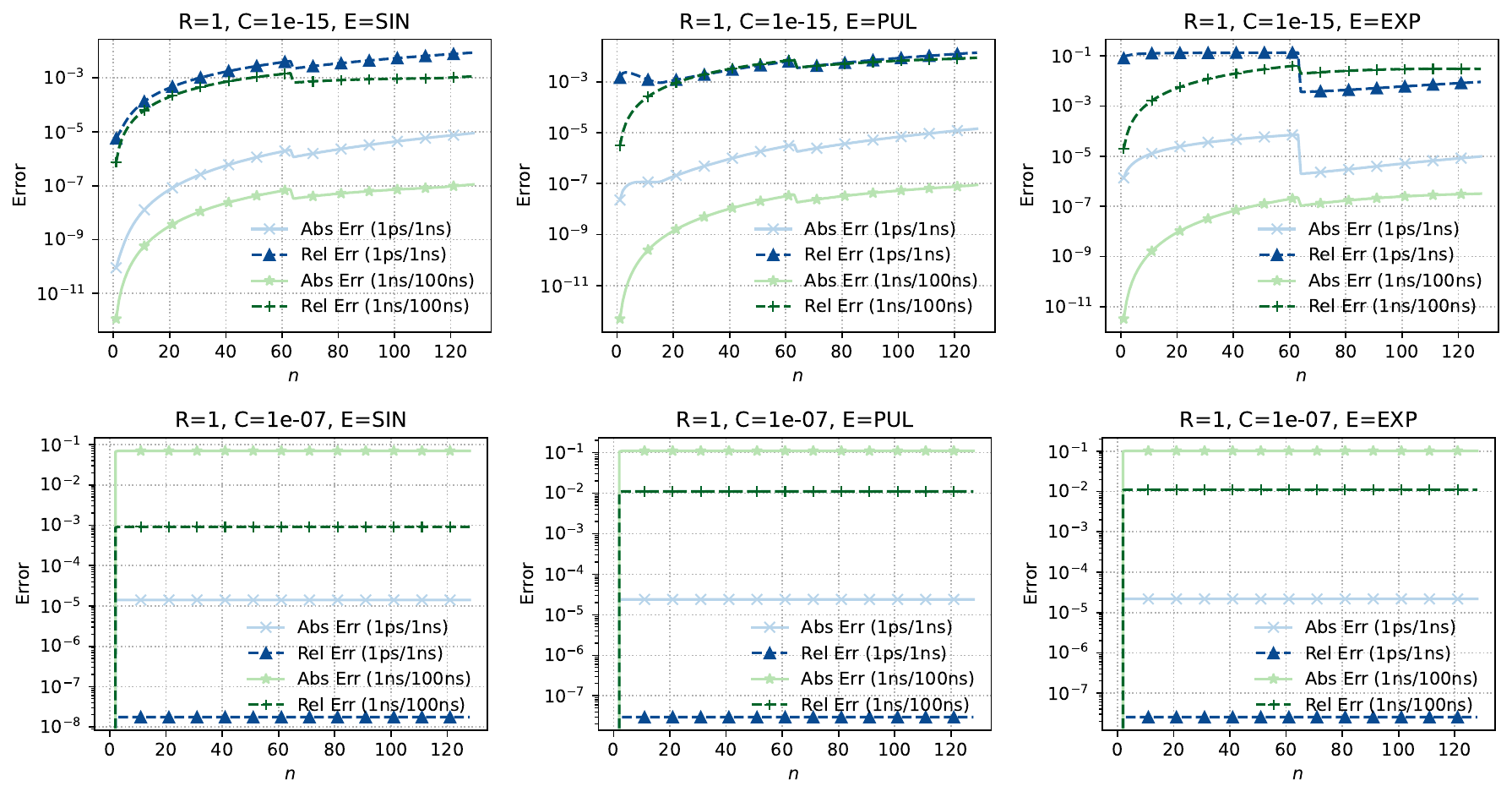}
    \caption{Absolute and Relative Error between \texttt{Ngspice} and Our Approaches on Function-Driven Simulations}
    \label{fig:combined_function}
\end{figure*}

To validate the correctness and measure the speedup of the proposed RC long-chain reduction criterion, we conducted two categories of experiments: \textbf{(1) Standard Function-Driven Simulations}; \textbf{(2) Practical Benchmark Tests}. The experiments mainly target circuits with relatively small time constant in the case \eqref{quick_node} and those with relatively large time constant in case \eqref{case3}. For the case \eqref{case2}, where the time constant is of the same order of magnitude as the simulation time ste, due to its complexity and the presence of multiple influencing factors, we defer a detailed investigation to future work. All experiments are conducted on an Arch Linux x86\_64 operating system (kernel version 6.12.8-arch1-1), and are performed using the open-source simulator \texttt{Ngspice} (version 44.2) \cite{ngspice} as the reference implementation. Our experiments code is \textbf{open-sourced} on Github\footnote{https://github.com/trrbivial/rc-chain-reduction-experiments/}.

\subsection{Function-Driven Simulations}

We generated \texttt{*.net} files representing the long-chain circuits as illustrated in Figure \ref{fig:chain_struct} for \texttt{Ngspice}. Each circuit file was parametrized by simulation step $s$, total simulation time $T$, resistance $R$, capacitance $C$, chain length $n$, and voltage input function $E$. The simulation output is the current $I(V_{\text{0}})$ during the time interval $[0,T]$. To ensure robustness, we configure \texttt{Ngspice} to use the \texttt{Sparse1.3} \cite{kundert1986sparse} solver for matrix computations.

We adopted simulation time steps of two orders of magnitude: $s/T = 1\text{ps}/1\text{ns}$ and $1\text{ns}/100\text{ns}$, corresponding to the time steps of $10^{-12}$ and $10^{-9}$ second, respectively. The resistance was fixed at $R = 1\Omega$, and two capacitance values were considered: $C = 10^{-15}$F (representing a relatively small time constant) and $C = 10^{-7}$F (representing a relatively large time constant). The \texttt{Ngspice} expressions of the voltage input functions $E$—common waveforms including \texttt{SIN}, \texttt{PULSE}, and \texttt{EXP}—are provided in Table \ref{tab:input_voltage_function}. The \texttt{SIN} functions are actually \texttt{COS} functions with initial voltage $1$V and at frequency 1GHz/100MHz; The \texttt{PULSE} fucntions will rise voltage from $-1$V to $1$V in 200ps/2ns and then drop back after the time point 500ps/50ns; The \texttt{EXP} functions will rise and drop voltage between $-4$V and $-1$V in an exponential way, please see the \texttt{Ngspice} Documentation \cite{ngspice} for detailed information.

\begin{table}[h]
\centering
\caption{Ngspice Expressions of Voltage Input Functions Used in Simulation}
\label{tab:input_voltage_function}
\begin{tabular}{ll}
\toprule
$\mathbf{s/T}$ & \textbf{Ngspice Expression (E)} \\
\midrule
1ps/1ns &  \texttt{SIN(0 1 1G 0 0 90)} \\
          &  \texttt{PULSE(-1 1 2PS 200PS 200PS 500PS 1NS)} \\
          &  \texttt{EXP(-4 -1 20PS 300PS 600PS 400PS)} \\
\midrule
1ns/100ns &  \texttt{SIN(0 1 100MEG 0 0 90)} \\
            &  \texttt{PULSE(-1 1 2NS 2NS 2NS 50NS 100NS)} \\
            &  \texttt{EXP(-4 -1 2NS 30NS 60NS 40NS)} \\
\bottomrule
\end{tabular}
\end{table}

When $C = 10^{-15}$F, we adopt the following strategy: if $n \leq 64$, we use equation \eqref{quick_node} for approximation; if $n > 64$, the chain is reduced to $m = \lfloor n/2 \rfloor$, and each grounded capacitor is scaled to $\frac{n+1}{m+1}C$. When $C = 10^{-7}$F, equation \eqref{case3} is employed for approximation.

We simulate every circuit using both \texttt{Ngspice} and our approach under various parameter settings. Each simulation yields a vector of current $I(V_0)$ of length $N$, denoted as $I_{\text{ref}}$ (\texttt{Ngspice} result) and $I_{\text{our}}$ (our result), respectively.

Direct computation of the mean of relative error can be numerically unstable when the absolute current values are very small (e.g., near 0), due to differences in discretization schemes between \texttt{Ngspice} and our method to calculate the current $I(V_0)$. To address this, We use the following weighted solution to calculate the mean of absolute and relative error:
\begin{align*}
    \Delta I_k = \left| I_{\text{ref}}[k] -I_{\text{our}}[k]\right|,\,\,\, &w_k = \left| I_{\text{ref}}[k]| +|I_{\text{our}}[k]\right|\\
     \displaystyle E_{\text{abs}} =\frac{1}{N}\sum_{k=0}^{N-1}\Delta I_k,\,\,\,\,\,\,\,\,\,\,\,\,\,\,\,\,\,\,  &
    \displaystyle E_{\text{rel}} = \left(\sum_{k=0}^{N-1}\Delta I_k\right)/\left( \sum_{k=0}^{N-1}w_i\right)
\end{align*}

As an illustrative example, consider two simulation time points where \texttt{Ngspice} outputs currents of $10^{-3}$A and $10^{-7}$A, while our method yields $1.01 \times 10^{-3}$A and $10^{-10}$A, respectively. Using the symmetric formulation $\Delta I_k / w_k$, the relative errors for the two points are $5 \times 10^{-3}$ and $1$, resulting in a misleading average of $0.5$. In contrast, our weighted method computes a relative error of $5.02 \times 10^{-3}$, which better captures the overall relative deviation.

The experimental results are summarized in Figure \ref{fig:combined_function}. When $C=10^{-15}$\,F, the absolute error remains negligible. For simulations using \texttt{SIN} and \texttt{PULSE} inputs with $n \leq 128$, the relative error is well-controlled within the range of $10^{-3}$ to $10^{-2}$. However, for \texttt{EXP} input under $n \leq 64$ and $s/T = 1\text{ps}/1\text{ns}$, the approximation using Equation~\eqref{quick_node} results in relative errors on the order of $10^{-1}$. For $n>64$, since we use the reduction method with $m = \lfloor n/2 \rfloor$, it yields better accuracy, and the relative error is below $10^{-2}$.

We further implemented a custom long-chain matrix solver to figure out why equation \eqref{quick_node} fails on \texttt{EXP} input. The results show that relative error profiles remain consistent with the observed discrepancy. We attribute this to differing discretization strategies for current calculation between \texttt{Ngspice} and ours, which is especially noticeable in circuits dominated by exponential dynamics. However, in terms of absolute errors, they still closely match those from \texttt{Ngspice}. Therefore, we think it's not a really serious problem.

Finally, when $C = 10^{-7}$\,F, the large time constant causes the circuit response to be dominated by the first two nodes of the RC chain. Subsequent nodes contribute negligibly to the overall current, resulting in flat error curves across all simulations.

\subsection{Benchmark Circuit Tests}

We use the original \texttt{*\_ann.net} files located in the \texttt{examples/klu} directory of the \texttt{Ngspice} distribution. These netlists are parsed to identify all eligible long-chain structures. For each such chain, we record the length $n$ and the starting node name $V_{\text{in}}$.

A simplified version \texttt{*\_simp.net} is then generated by deleting all nodes and edges in the chain except for $V_{\text{in}}$, and replacing the grounded capacitor at $V_{\text{in}}$ with $(n+1)C$. As for simulation results, the original \texttt{*\_ann.net} files contain instructions for  the output of the voltage at several nodes, for example, the command:  \texttt{wrdata output.dat V(g1339\_1)}. If an output node is preserved in the simplified circuit, the corresponding command remains unchanged. However, if an output node such as \texttt{g1339\_1} is removed (e.g., it lies within a reduced chain), and the chain starts at \texttt{netg1339\_1\_1}, then in \texttt{*\_simp.net}, we replace the output expression \texttt{V(g1339\_1)} with \texttt{V(netg1339\_1\_1)}, while the corresponding  \texttt{*\_ann.net} file is left unmodified throughout.

\begin{table}
  \centering
  \caption{Environment Settings for Benchmark Circuit Tests}
  \label{tab:experiment_settings}
  \begin{tabular}{ll}
    \toprule
    \textbf{Environment} & \textbf{Settings}\\
    \midrule
    OS & Arch Linux x86\_64 (6.12.8-arch1-1) \\
    PC & 82JD Lenovo Legion Y9000P2021H \\
    CPU & 11th Gen Intel i7-11800H (16) @ 4.600GHz\\
    MEM (Byte) & L1(32K) L2(1.25M) L3(24M) DRAM(32G) \\
    \texttt{Ngspice} (44.2) & \texttt{GMIN=1e-015 ABSTOL=1e-13 ACCT KLU}\\
    TESTs & \texttt{warmup=3, nrun=10, using geomean} \\
  \bottomrule
\end{tabular}
\end{table}

\begin{figure}[h]
  \centering
  \includegraphics[width=\linewidth]{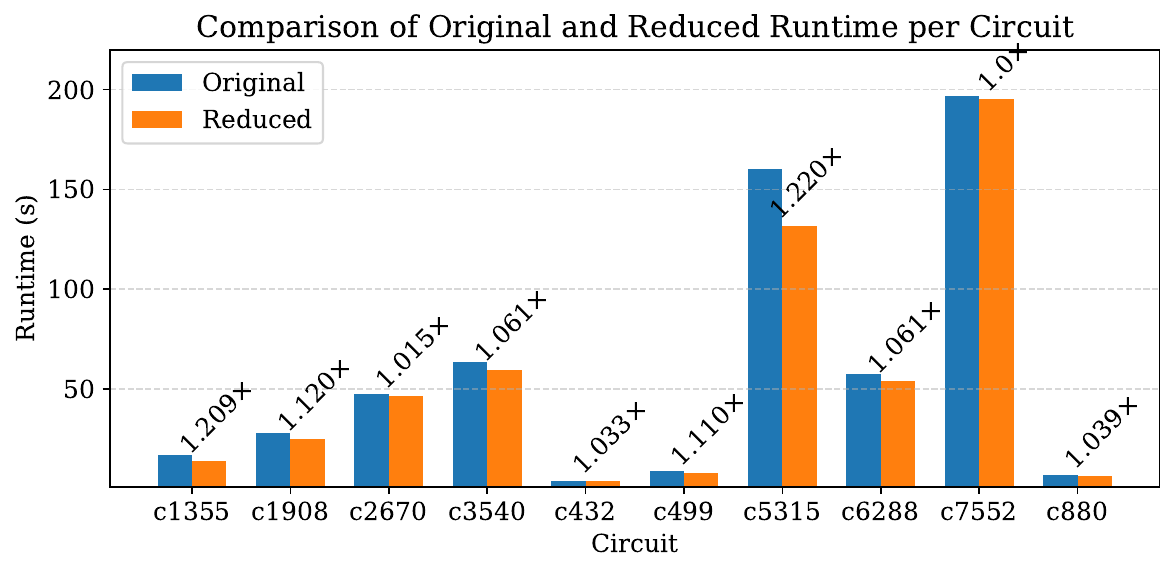}
  \caption{Geometric Mean of Tested Runtimes and the Speedup Our Reduction Achieved Compared to the Original on Benchmark Circuits.}
  \label{fig:speedup_of_reduction}
\end{figure}

Finally, we simulate both \texttt{*\_ann.net} and \texttt{*\_simp.net} using \texttt{Ngspice} and compare the resulting outputs, fetch the total transient simulation time in seconds provided by \texttt{Ngspice} log file when option \texttt{ACCT} is set. To improve simulation efficiency, \texttt{Ngspice} is configured to use the \texttt{KLU} \cite{davis2010algorithm} solver for matrix computations. Table \ref{tab:experiment_settings} summarizes the environment configuration used for benchmark circuit tests. Table \ref{tab:final_exp_abs_rel} shows the absolute and relative error between \texttt{Ngspice} and our approach on each benchmark circuit. Figure \ref{fig:speedup_of_reduction} shows the geometric mean of tested runtime and the speedup we achieved compared to the original.

Experimental results demonstrate that our method achieves an average performance improvement of \textbf{8.8\%}, while maintaining a relative error below \textbf{0.7\%}. Notably, significant speedups are observed on circuits \texttt{c1355} and \texttt{c5315}, with improvements of \textbf{21\%} and \textbf{22\%} respectively. Additionally, circuits \texttt{c1908} and \texttt{c499} exhibit performance gains exceeding 10\%, while \texttt{c3540} and \texttt{c6288} achieve approximately 6\% improvement.

However, for the remaining circuits, the performance gains are relatively modest, mostly below 4\%, and no improvement is observed for circuit \texttt{c7552} for sure because it has no long-chain structure.

\begin{table}
    \centering
    \caption{Absolute and Relative Error between \texttt{Ngspice} and Our Approach on Benchmark Circuits}
    \begin{tabular}{lcc}
    \toprule
    \textbf{Circuit} & \textbf{AbsErr (Mean)} & \textbf{RelErr (Mean)} \\
    \midrule
    \texttt{c1355} & 5.3282e-04 & 3.5562e-03 \\
    \texttt{c1908} & 1.4244e-03 & 4.0484e-03 \\
    \texttt{c2670} & 4.8238e-04 & 1.7857e-03 \\
    \texttt{c3540} & 0.0000e+00 & 0.0000e+00 \\
    \texttt{c432}  & 2.6064e-07 & 8.6192e-06 \\
    \texttt{c499}  & 8.9443e-04 & 2.1560e-03 \\
    \texttt{c5315} & 1.5915e-03 & 6.0819e-03 \\
    \texttt{c6288} & 2.3137e-05 & 1.0502e-04 \\
    \texttt{c7552} & 0.0000e+00 & 0.0000e+00 \\
    \texttt{c880}  & 2.0711e-04 & 1.2940e-03 \\
    \bottomrule
    \end{tabular}
    \label{tab:final_exp_abs_rel}
\end{table}

\section{Related Works}

\textbf{TICER-based Reduction and Extensions.}
TICER~\cite{810649, 4271561} provides a realizable method for reducing RC circuits through time-constant balancing, and has inspired a series of extensions. HD-TICER~\cite{9525071} extends TICER to high-dimensional RC block elimination, while CGAT-TICER~\cite{cgatticer} leverages graph attention networks to improve node selection. P-TICER~\cite{10617546} introduces a parallel acceleration framework for TICER, and GNN-based TICER~\cite{9963409} further integrates neural network-based reduction strategies.

\textbf{Graph Sparsification and Spectral Methods.}
Several works adopt graph-theoretic and spectral sparsification techniques for scalable model order reduction. Effective resistance-based sparsification~\cite{doi:10.1137/080734029, DBLP:journals/corr/abs-0808-4134} and spectral methods~\cite{7001355, 9358096, zyb2024rcreduction} have been shown effective in preserving circuit accuracy while reducing complexity. Matrix sparsification for dense RLC systems~\cite{8806988, 10.1145/3287624.3287658} enables efficient simulation of large networks.

\textbf{Classic and Structure-Preserving Model Reduction.}
Traditional reduction techniques such as PRIMA~\cite{712097} and TurboMOR-RC~\cite{7410004} offer passive, structure-preserving models, while SIP~\cite{4681658} and RLCSYN~\cite{4253237} generalize to many-terminal or RLC interconnects. Optimal elimination frameworks~\cite{20121117} further improve solution accuracy in large RC networks.

\textbf{Sparse Matrix Solvers.}
KLU \cite{davis2010algorithm} represent prominent sparse matrix solver for circuit simulations.  SparseLU-FPGA \cite{splufpga} is a hardware implemtation of the Sparse LU factorization. Spatula~\cite{10.1145/3613424.3623783} is a state-of-the-art hardware accelerator for sparse matrix factorization. General sparse matrix techniques are also discussed in~\cite{sipics2007}.

\textbf{Hardware-Accelerated Circuit Simulation.} FPGA-based solvers for SPICE simulation \cite{kapre2009parallelizing} offer high-throughput alternatives.

\textbf{Parallel and Distributed Circuit Simulation.}
Efficient parallel simulators such as Xyce~\cite{doecode_2462, osti_771528} support multi-threaded circuit evaluation, while domain decomposition~\cite{4796513} and multicore scheduling~\cite{chen2012parallel, benk2017holistic} enable scalable transient simulation. Classical parallel solvers~\cite{https://doi.org/10.1002/1099-1506(200010/12)7:7/8<649::AID-NLA217>3.0.CO;2-W} also provide foundations for large-scale matrix solving.

\textbf{Open-Source and Industrial Simulation Tools.}
Several commercial circuit simulators remain widely used in both industry and academia, including \texttt{HSPICE}~\cite{hspice}, \texttt{Spectre}~\cite{spectre}, \texttt{Eldo}~\cite{eldo}, and the open-source \texttt{Ngspice}~\cite{ngspice}. Additionally, Qucs~\cite{10.5555/1568533.1568534} provides a GPL-licensed simulation platform supporting compact device models.

\section{Conclusion}

In this work, we proposed \textbf{three} reduction strategies tailored to different scales of time constants for accelerating transistor-level circuit simulation.

Experimental results demonstrate that our method yields an average performance improvement of \textbf{8.8\%} (up to \textbf{22\%}) on simulating benchmark circuits which include a variety of functional modules such as ALUs, adders, multipliers, SEC/DED checkers, and interrupt controllers, with only \textbf{0.7\%} relative error.


\bibliographystyle{ACM-Reference-Format}
\bibliography{sample-base}

\end{document}